\documentclass[twocolumn,reprint,superscriptaddress,amsmath,amssymb,aps,nofootinbib]{revtex4-1}

\usepackage{graphicx}
\usepackage{dcolumn}
\usepackage{bm}

\usepackage{hyperref}
\usepackage[utf8]{inputenc}
\usepackage[dvipsnames]{xcolor}
\usepackage[normalem]{ulem}
\usepackage{xspace}
\usepackage{fontawesome} 
\usepackage{multirow}

\usepackage[dvipsnames]{xcolor}
\usepackage[utf8]{inputenc}
\hypersetup{
    colorlinks=true,
    citecolor = blue,
    linkcolor=red,
    filecolor=magenta,      
    urlcolor=magenta,
}

\newcommand{\NAP}{Dipartimento di Fisica "Ettore Pancini", Università degli studi di Napoli ``Federico II''; and INFN - Sezione di Napoli. Complesso Univ. Monte S. Angelo, I-80126 Napoli, Italy}
\newcommand{\NBIA}{Niels Bohr International Academy, Niels Bohr Institute, University of Copenhagen, Copenhagen, Denmark}

\begin{document}
\title{Axions from Neutron Star Mergers}

\author{Damiano F. G. Fiorillo}
\affiliation{\NAP}
\affiliation{\NBIA}
\author{Fabio Iocco}
\affiliation{\NAP}

\date{\today}

\begin{abstract}
Axion-like particles (ALP) can in principle be produced in very hot and dense astrophysical environments, escape from the extreme object where such conditions are met, and then be converted in gamma--rays in the magnetic fields intervening between the event and the Earth. This process potentially offers a new window on both the physics of the axions, and the inner working of the astrophysical objects where they are produced. 
Interestingly, while this process has been studied for core--collapse supernovae and other extreme astrophysical events, no estimate exists for Neutron Star Mergers, objects recently identified through the detection of gravitational waves. 
 
In this work we study the production of ALPs in neutron star mergers, finding that for a large region of the ALP parameter space its magnitude at the source is such to produce a sizable gamma-ray signal at Earth.
We show detection forecasts for such events placed in nearby galaxies, finding that they are potentially observable with the Fermi--LAT, thus opening a new window into both the astrophysics of these cataclysmic events, and of new particles beyond the standard model.

\end{abstract}


\maketitle

\section{Introduction}
\par Axion-like particles (ALPs) can in principle be produced in the peculiarly hot and dense environments of extreme astrophysical objects. This process has been studied for magnetars~\cite{Fortin:2021sst,Lloyd:2020vzs}, pulsars~\cite{Archer:2020znv}, neutron stars~\cite{Ganguly:2013bca}, compact binaries~\cite{Kavic:2019cgk}, black holes~\cite{Cardoso:2018tly}, gamma-ray bursts~\cite{Mena:2011xj,Covino:2016cuw,Berezhiani:1999qh,Reynoso:2017mal}, active galactic nuclei~\cite{Ayad:2019hrj, Mena:2013baa,Horns:2012pp,DeAngelis:2011id}, and supernovae (SNe)~\cite{Raffelt:1996wa,Raffelt:2006cw,Fischer:2016cyd,Turner:1987by,Burrows:1988ah,Burrows:1990pk,Raffelt:1987yt,Raffelt:1990yz,Keil:1996ju,Hanhart:2000ae,Chang:2018rso,Carenza:2019pxu,Grifols:1996id,Brockway:1996yr,Payez:2014xsa,Jaeckel:2017tud,Meyer:2016wrm,Meyer:2020vzy,Calore:2020tjw}.

The actual observation of a neutron star merger (NSM) through gravitational waves \cite{TheLIGOScientific:2017qsa} produced in the event has prompted the exploration of a realm of extreme densities and temperatures, 
as addressed in \cite{Radice:2018pdn,Radice:2016rys,Wanajo:2014wha,Kasen:2013xka,Metzger:2017wot,Baiotti:2016qnr,Shibata:2006nm,Baiotti:2008ra,Hotokezaka:2011dh}. 

The physical conditions achieved in such extreme environment are comparable with those achieved in some of the objects listed above, and it therefore makes sense to address the study of ALPs in NSMs. 
In such unique environment, the bremsstrahlung process of neutrons $N+N\to N+N+a$ could be an efficient way of producing ALPs, a process that has been studied in the context of core--collapse SNe~\cite{Turner:1988bt,Carena:1988kr,Iwamoto:1992jp,Brinkmann:1988vi,Keil:1996ju,Carenza:2019pxu}, leading to constraints on the ALPs properties from the non-observation from the SN1987A~\cite{Raffelt:1996wa,Graham:2015ouw,Burrows:1988ah,Mayle:1989yx,Raffelt:1987yt,Turner:1987by}, and to forecasts on future SN detections~\cite{Jaeckel:2017tud,Meyer:2016wrm,Calore:2020tjw}.

The possibility that ALP production could take place in NSMs has been addressed in~\cite{Dietrich:2019shr, Harris:2020qim}, where the authors address the potential cooling of the object, caused by ALPs streaming through the dense material of the merger, whose structure could hence be modified.

Here we address a complementary question: whether the flux of ALPs generated in the whole merger --within the limitations arising from  current observations of gravitational waves-- is such to be potentially observable at Earth.
Namely whether there is a region of the ALP parameter space --still allowed by observations-- where ALP emission from NSM is sizable, and if so whether any observational signature would be left by ALPs being converted into photons in the magnetic field intervening between us and the merger. In Section \ref{sec:constraints} we independently identify the region of ALP parameter space which is unconstrained by present SN observations, shown in Fig.~\ref{fig:photonconstraints}. 

In Section \ref{sec:earthflux} we comment on our main finding, namely that for a sizable fraction of this region 
the merger structure is preserved, 
and a signal in soft gamma--rays is potentially observable with the Fermi-LAT detector~\cite{Fermi-LAT:2009ihh},
thus offering the possibility of a simultaneous identification with the gravitational waves produced in the merger.
This opens a new exciting window for physics beyond the standard model, and in the multimessenger study of NSMs.

\section{ALP fluxes from Neutron Star mergers}\label{sec:axionfluxes}

\par We consider ALPs produced via nucleon-nucleon ALP bremsstrahlung, a process has been thoroughly studied in the literature~\cite{Brinkmann:1988vi,Iwamoto:1992jp,Carenza:2019pxu,Harris:2020qim}. Since strong interactions in dense nucleon environments are poorly known, they are often treated at the level of one-pion exchange (OPE): with this approach, Ref.~\cite{Harris:2020qim} determines the ALP emissivity in the hypermassive neutron star which forms for a short time at the end of a NSM. Recently, Ref.~\cite{Carenza:2019pxu} has shown that going beyond the OPE approximation can lead to a reduction by roughly an order of magnitude in the ALP emissivity. For this work we adopt the emissivity from~\cite{Harris:2020qim}, noting that this procedure does not change the qualitative conclusions of our work nor the implications on detectability, as discussed in detail in Sec.~\ref{sec:constraints}.  
We use the emissivity for the full phase-space calculation for non-relativistic neutrons, assuming a density equal to the saturation density of neutrons. For relativistic neutrons and for higher densities the emissivity would be larger, our choice is thus conservative.

Furthermore, Ref.~\cite{Harris:2020qim} only considers the ALP production due to neutron-neutron ALP bremsstrahlung. As suggested in Ref.~\cite{Fore:2019wib}, neutron stars could also contain a population of protons and pions. An analogous situation for SN leads to an enhanced axion emissivity via the process $\pi^-p\to na$, ~\cite{Carenza:2020cis}. This component would also be harder, with ALPs produced mostly with hundreds of MeV. Furthermore, in the presence of protons, neutron-proton ALP bremsstrahlung would also contribute to the emissivity. For this work, we conservatively do not consider these components, because of the large uncertainties in the concentrations of pions and protons within the NSM. 

Following Ref.~\cite{Harris:2020qim}, we define the interaction Lagrangian for neutrons and ALPs as
\begin{equation}
    \mathcal{L}_\text{aN}=G_\text{aN} \partial_\mu a \bar{N}\gamma^\mu \gamma_5 N,
\end{equation}
where $G_\text{aN}$ is the coupling constant. For future convenience, we parameterize the ALP emissivity (measured in erg cm$^{-3}$ s$^{-1}$) as a function of the temperature $T$ as $Q(T)=\left(\frac{G_\text{aN}}{G_\text{SN}}\right)^2 Q_\text{SN} (T)$, where $G_\text{SN}=7.8\times 10^{-10} \text{GeV}^{-1}$ and $Q_\text{SN} (T)$ is extracted from Fig.~7 of~\cite{Harris:2020qim}; $G_\text{SN}$ is the maximum coupling allowed by the observation of SN1987A based on energy-loss arguments.

\begin{figure}
    \centering
    \includegraphics[width=0.4\textwidth]{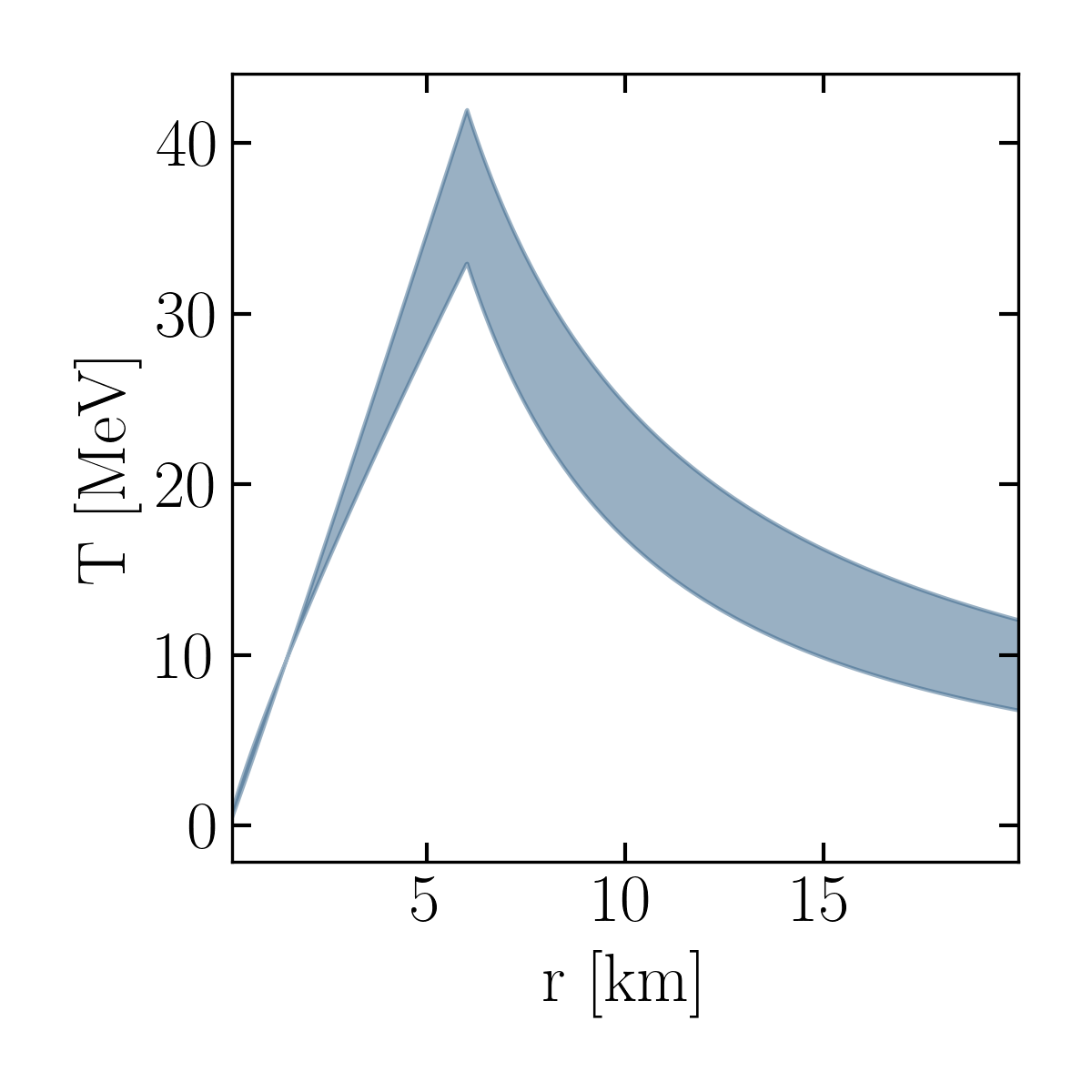}
    \caption{\textbf{Temperature profile of the NSM.} We show the band of spherically symmetric temperature profiles adopted for our calculation as a function of the radius. The profiles have been obtained as a broken power-law spectrum adapted from those shown in Fig. 4, left panel of~\cite{Hanauske:2019qgs}.}
    \label{fig:temperatureprofile}
\end{figure}

The ALPs produced are assumed to freely escape, as shown in~\cite{Harris:2020qim}. For their spectral distribution we follow the Fermi surface approximation of~\cite{Ishizuka:1989ts} in the degenerate neutrons limit, which gives a slightly modified blackbody spectrum. We have also tested a simpler blackbody emission finding no significant differences in the final result.
For the total number of ALPs emitted by the source we normalize this spectrum to the emissivity $Q(T)$ and integrate over the volume of the source. The temperature profile of the object is quite uncertain. The curve we use, shown in Fig.~\ref{fig:temperatureprofile}, is a spherically symmetric broken power-law, adapted to reproduce the order of magnitude of the profile in Fig.~4, left panel of Ref.~\cite{Hanauske:2019qgs}.
The two profiles roughly bracket the upper and lower temperature envelope to the order of magnitude. This profile was obtained by a simulation after $25$ ms from the start of the merger. At shorter times even larger temperatures could be reached (as shown in Fig.~3 of~\cite{Hanauske:2019qgs}).

The duration of the event is also quite uncertain. Studies on the NSM connected with the gravitational wave GW170817 conclude that the supermassive neutron star at the center survived for about 1~s~\cite{Gill:2019bvq,Harris:2020qim}, and simulations report the temperature profiles for times up to about 25~ms after the burst beginning. Here we assume a duration of 1~s. 
The duration of the burst may actually be shorter, and/or the temperature profile we have adopted --and on which the emissivity depends crucially-- may not hold for the entire duration of the burst. This would of course affect our conclusions: a duration of the burst of only 10~ms (100 times shorter than the one shown in the plots) would make only a burst in the LMC visible with Fermi-LAT. It is however unlikely that all conditions above would conjure for reducing the flux as much, also given that our other assumptions are conservative: in particular, we are using a constant density for the nuclear matter, whereas it might reach larger values near the center of the hypermassive neutron star. Therefore, we present our results according to this choice, which was also similarly reported in Ref.~\cite{Diamond:2021ekg}.

With $G_{\text{aN}}=1\times10^{10}$ GeV$^{-1}$, the total energy injected for an event duration of 1~s into ALPs varies between $3\times10^{51}$ erg to $1\times10^{52}$ erg for the two temperature profiles adopted. For comparison, the energy radiated in gravitational waves was estimated in~\cite{TheLIGOScientific:2017qsa} to be larger than $4\times10^{52}$ erg.

\section{Constraints on ALP parameter space} \label{sec:constraints}
\begin{figure}[t!]
    \centering
    \includegraphics[width=0.5\textwidth]{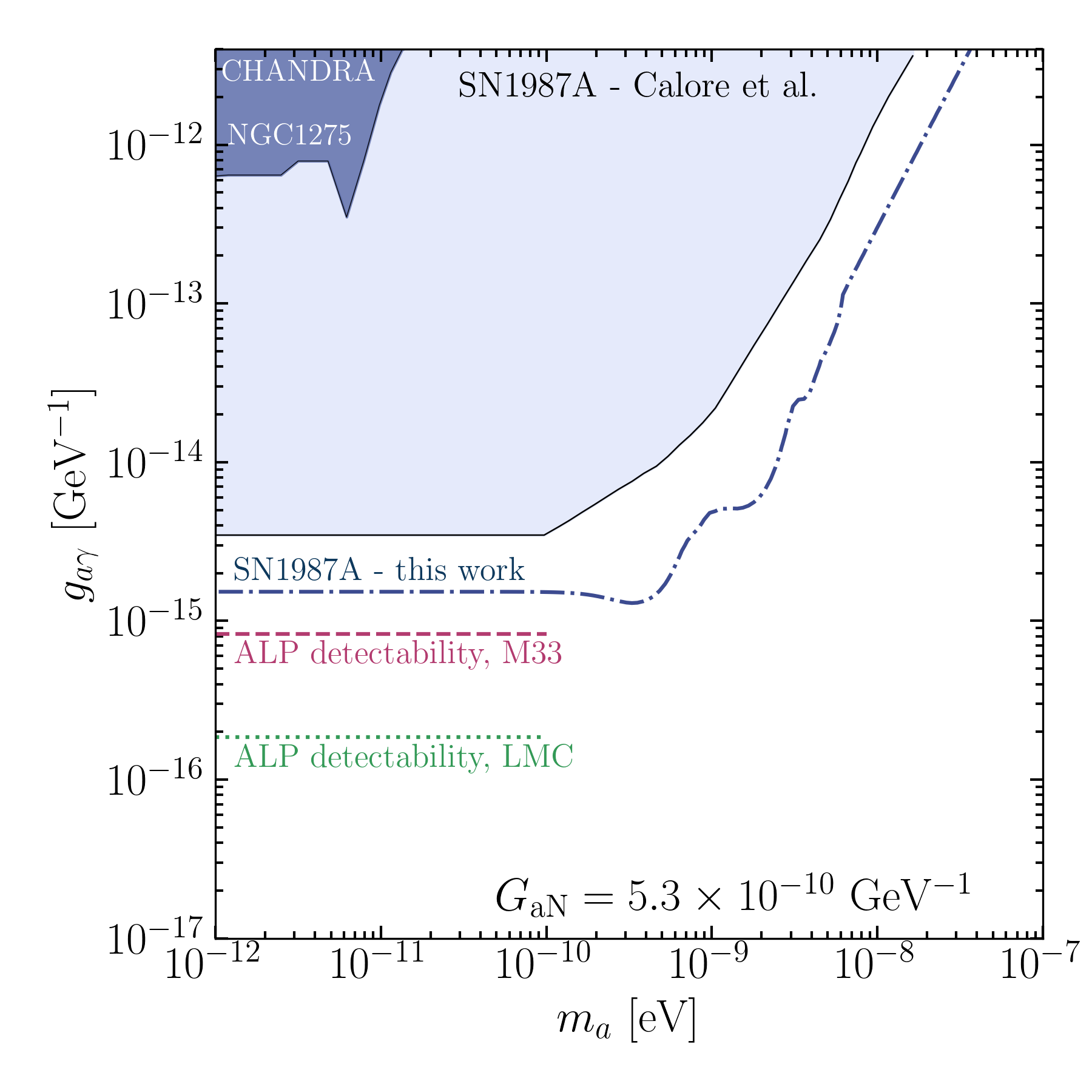}
    \caption{\textbf{Projected constraints on ALP parameter space by future observations of SN.} 
    Constraints from the non-observation of gamma-rays from ALP conversion from SN1987A, from literature and as independently obtained for self-consistency in this work. We choose the value of $G_{\text{aN}}=5.3\times10^{-10}$~GeV$^{-1}$. For the same reference value of $G_{\text{aN}}$, we also show the threshold values of $g_{a\gamma}$ leading to a detectable gamma-ray signal in a NSM happening in M33 (dashed) and LMC (dotted), for $m_a<10^{-10}$~eV (see Sec.~\ref{sec:earthflux}). }
    \label{fig:photonconstraints}
\end{figure}

Before showing our results on the gamma-ray fluxes expected from a NSM, in this section we report the present constraints on ALP properties from SN1987A. Be noticed that similar constraints have been derived also elsewhere, here we independently and self-consistently obtain them within our setup, and report them for reference and consistency. This allows us to show that the expected gamma-ray flux from nearby NSMs is sizable, for a wide region of the parameter space as yet permitted by observations.

Constraints from the non-observation of gamma-rays by EGRET at the time of SN1987A have been drawn on the coupling $g_{a\gamma}$~\cite{Brockway:1996yr,Grifols:1996id,Payez:2014xsa}. These depend only on the Primakoff production of ALPs. Here we focus on ALP production via nucleon bremsstrahlung, which can only constrain the product $g_{a\gamma} G_{\text{aN}}$, since $G_{\text{aN}}$ determines the rate of ALP production and $g_{a\gamma}$ determines the conversion in galaxy. This approach is followed in Ref.~\cite{Calore:2020tjw}, where the constraints on $g_{a\gamma}$ are determined for a fixed value of $G_{\text{aN}}=5.33\times10^{-10}$ GeV$^{-1}$
\footnote{Before comparing with Ref.~\cite{Calore:2020tjw}, we note that the emissivity we adopt here differs from that used in Refs.~\cite{Calore:2020tjw,Carenza:2019pxu}. As mentioned in Sec.~\ref{sec:axionfluxes}, this difference 
does not affect our conclusions on the detectability. In fact, the normalization of the ALP flux depends on the coupling $G_{an}$, for which the main constraints come from the ALP emission in SN1987A (either because of cooling or because of ALP to gamma conversion). Therefore a discrepancy in the emissivity is reabsorbed in a redefinition of the coupling $G_{an}$: in other words, if the emissivity were 10 times lower, the constraints on $G_{an}$ would be a factor of $3$ weaker, and therefore the signal we obtain would still be allowed by the present parameter space.}.

To ensure consistency, we independently compute the $95\%$ constraints on $g_{a\gamma}$ from the non-observation of gamma-rays from SN1987A by EGRET, requiring that the photon fluence in the interval from 25 MeV to 100 MeV from SN1987A was smaller than $0.6$ cm$^{-2}$~\cite{Payez:2014xsa}. We show the corresponding constraint in Fig.~\ref{fig:photonconstraints} as a dot-dashed line, to be compared with the cyan exclusion region obtained in Ref.~\cite{Calore:2020tjw}. The blue curve is stronger by about a factor of $3$: since the energy fluence is proportional to $g_{a\gamma}^2$, this implies a factor of 10 difference in the gamma-ray fluxes, which coincides with what we expected for the different emissivities adopted.

We also show in the same figure the threshold values of $g_{a\gamma}$ (for the reference value $G_{\mathrm{aN}}=5.3\times 10^{-10}$~GeV$^{-1}$) which would lead to a detection of the gamma-ray signal from ALP conversion at Fermi-LAT for a NSM happening in the near galaxies M33 and LMC. These are obtained by requiring the minimum gamma-ray flux determined in Sec.~\ref{sec:earthflux} to be above the Fermi-LAT sensitivity. Since the gamma-ray flux is determined by the product $g_{a\gamma}G_{\mathrm{aN}}$, a different choice of $G_{\mathrm{aN}}$ would simply correspond to a proportional scaling of the limits in Fig.~\ref{fig:photonconstraints}. 


\section{ALP from NSM: Gamma--ray flux at Earth} \label{sec:earthflux}

\begin{figure}[t!]
    \centering
    \includegraphics[width=0.4\textwidth]{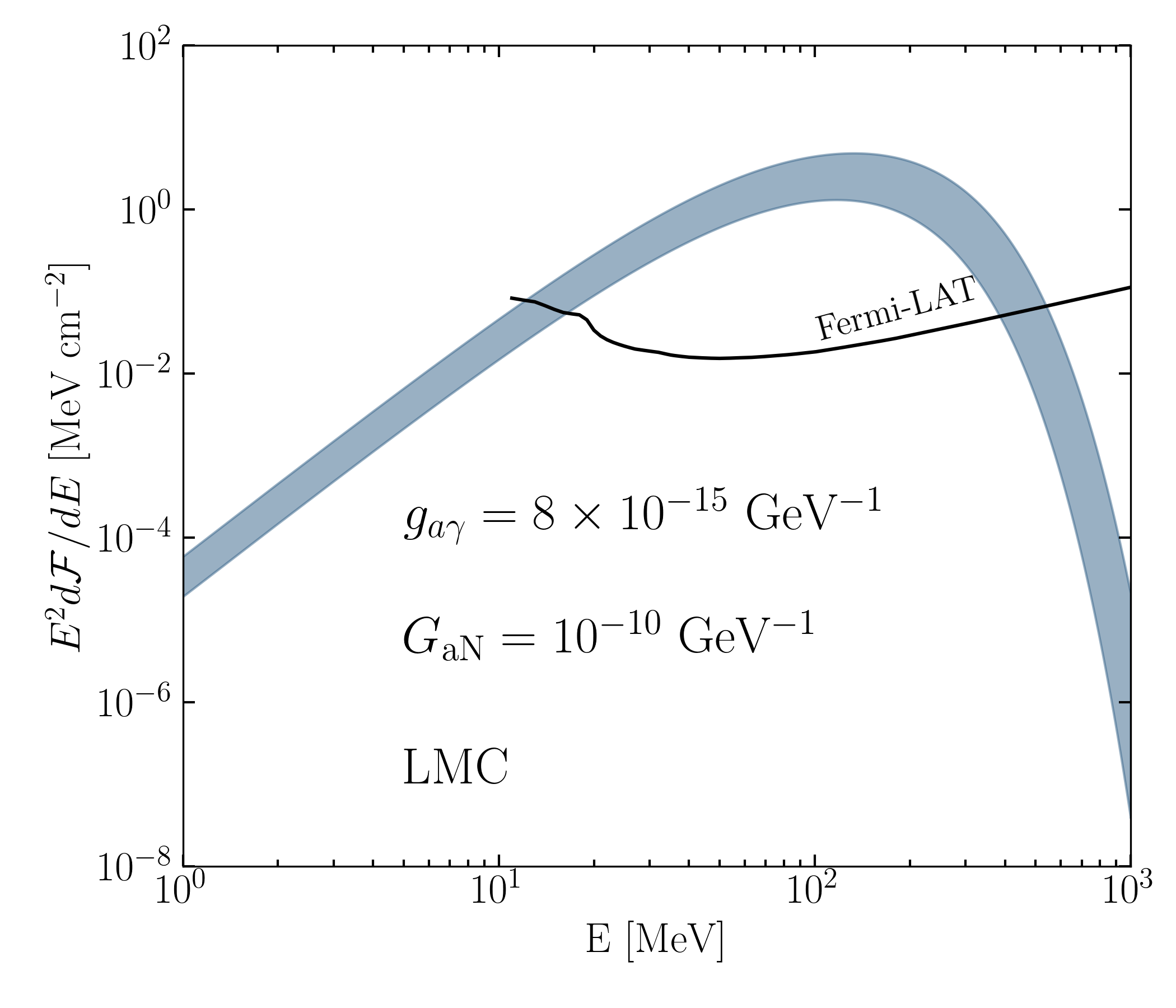}
    \includegraphics[width=0.4\textwidth]{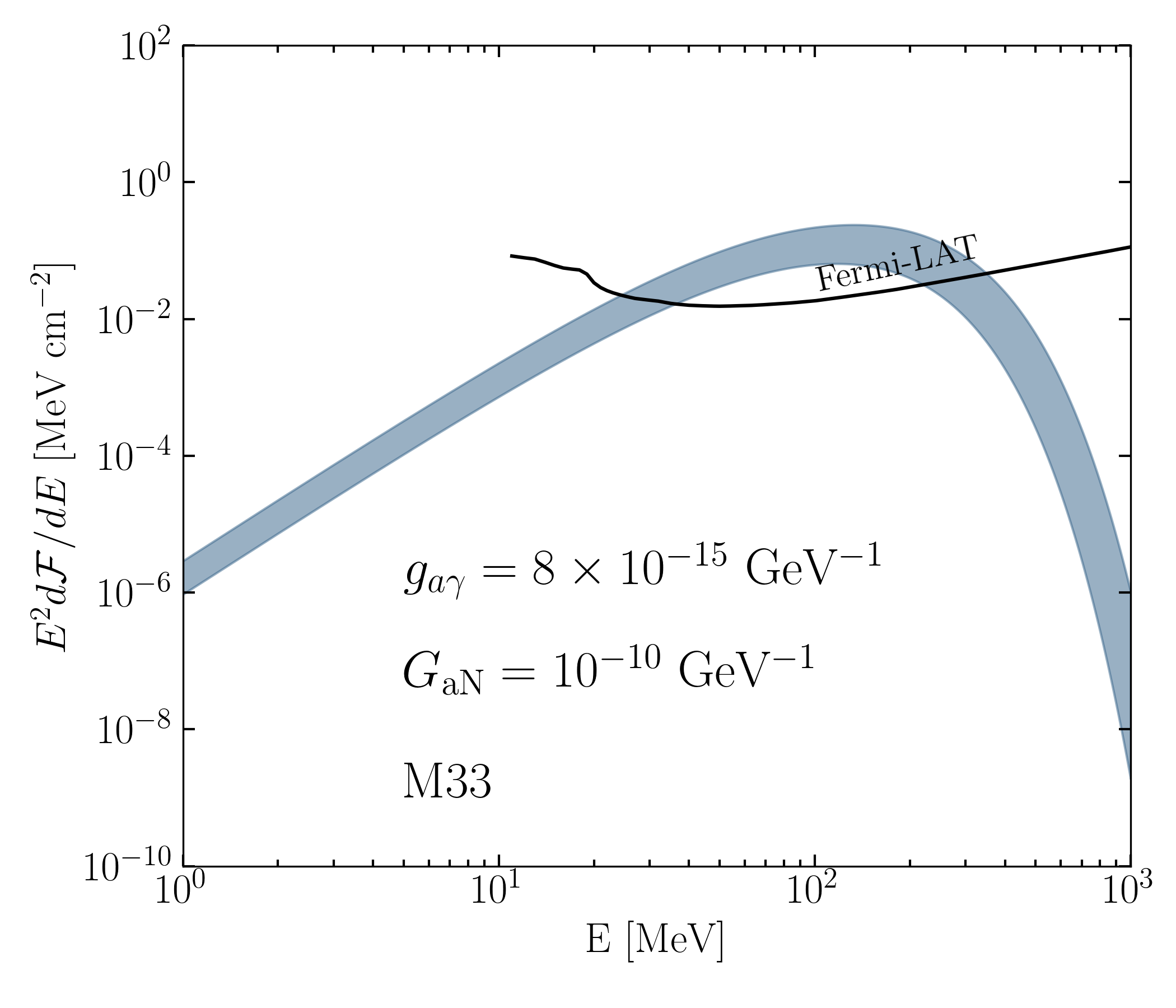}
    \caption{\textbf{Gamma-ray fluence from conversion of ALPs produced in the NSM.} We show the gamma-ray fluence of a NSM, obtained for the band of temperature profiles in Fig. \ref{fig:temperatureprofile} for a benchmark choice of ALP parameters shown in the text. We consider the event happening in the LMC (top), as in the case of SN1987A, and in M33 (bottom). We show in black the 90\% sensitivity of the Fermi-LAT, obtained from the effective area in Ref.\cite{Fermiarea} and assuming no background (for exposure times of 10~s the background rate provided in the Table leads to less than 1 event). The ALP mass is chosen as smaller than $10^{-10}$~eV, see text for details. We require 2.44 events per energy decade for observation in the background-free regime~\cite{Feldman:1997qc}.
    }
    \label{fig:photonflux}
\end{figure}
In~\cite{Harris:2020qim} the effects of the cooling via ALP production on the neutron star merger were emphasized. A complementary possibility for detectability, is the observation of the ALPs after their conversion into gamma-rays in the galactic magnetic field (we conservatively neglect the intergalactic magnetic fields, whose strength is largely uncertain). 

During the propagation in the Milky Way ALPs can convert into gamma-rays via two-photon coupling with the galactic magnetic field. The interaction term in the Lagrangian is
\begin{equation}
    \mathcal{L}_{a\gamma}=-\frac{1}{4} g_{a\gamma} F_{\mu\nu} \tilde{F}^{\mu\nu} a,
\end{equation}
where $a$ is the ALP field, $F_{\mu\nu}$ is the electromagnetic field tensor, $\tilde{F}^{\mu\nu}$ is the dual field tensor, and $g_{a\gamma}$ is the coupling constant. To estimate the probability of conversion of ALPs to gamma-rays $P_{a\to\gamma}(E,b,l)$ (as a function of the ALP energy $E$, and of the galactic coordinates $b$ and $l$ identifying the chosen line of sight), we solve perturbatively the transport equations in Ref.~\cite{Calore:2020tjw}, using the Jansson-Farrar 12c model~\cite{Jansson:2012pc,Adam:2016bgn} for the galactic magnetic field.

The gamma-ray fluence emitted over the entire duration of the event is shown in Fig.~\ref{fig:photonflux} (top panel) for the NSM, with a temperature profile taken from Ref.~\cite{Carenza:2019pxu}. 
The ALP couplings $G_{\text{aN}}$ and $g_{a\gamma}$ have been chosen so as to saturate the bounds from SN1987A observations by EGRET shown in Fig.~\ref{fig:photonconstraints}. The result is insensitive to the ALP mass when the latter is chosen to be sufficiently small, in particular when $m_a\ll 10^{-10}$~eV, which is the regime in which Fig.~\ref{fig:photonflux} has been obtained. With this choice, a NSM in LMC, the same position as SN1987A, would give rise to a signal detectable by Fermi-LAT. For a NSM in the M33 galaxy, with a distance of more than one order of magnitude larger than LMC, the fluxes are correspondingly suppressed. Nonetheless, the predicted signal is still within the reach of Fermi-LAT, indicating that even accounting for the uncertainties on the temperature profile does not jeopardize a possible detection. 

We do not include any background radiation produced by the NSM. 
Actually, NSMs have long been associated with short gamma-ray bursts, whose radiation can indeed be prominent in the hundreds of MeV range. However, the significant case of GW170817 led to a gamma-ray burst detected almost two seconds later than the merger~\cite{Veres:2018trt}. Therefore, the two signals could potentially be distinguished on temporal basis. In this regard, we mention that the timing of the neutron star merger for ALP detection would be entirely determined by the gravitational wave signal, whose properties allow to identify the instant of formation of the hypermassive neutron star from the merger.

Finally, we comment on the observability of the diffuse gamma--rays emitted by the whole population of NSMs. The rate of NSMs is much lower than that of core-collapse SNe, so that the diffuse production associated to the NSM population is much lower than that of SNs. The ALP--induced diffuse gamma--ray production from a SN population has been studied in Ref.~\cite{Calore:2020tjw}, where it is shown that the constraints from individual SN1987A are much stronger than the ones from diffuse production. We have explicitly checked that this is true for diffuse production from NSMs, due to their lower rate, as from the literature~\cite{LIGOScientific:2016hpm,TheLIGOScientific:2017qsa,Pian:2020vul,Ciolfi:2020huo}.

\section{Conclusions}\label{sec:conclusions}
\par High-energy astrophysical events are progressively expanding our opportunities to explore physics beyond the Standard Model. In this analysis, we focus on Neutron Star Mergers as a source of Axion Like Particles potentially detectable at Earth after their conversion in gamma-rays in the intervening magnetic field. We focus on the nucleon-nucleon ALP bremsstrahlung mechanism for the production of ALPs in NSMs, a possibility that has not been previously discussed in the literature\footnote{While this paper was in preparation, Ref.~\cite{Diamond:2021ekg} appeared, discussing an idea similar to ours: namely that dark photons may be produced in neutron star mergers and be detected at Earth.
Whereas we are glad to notice that their ``recipes'' for the local physics (NSM duration, temperature and density profile), and that their conclusions about the detectability are similar to ours, our paper focuses on an entirely different production mechanism via ALPs rather than dark photons.} 
\\
We find that Neutron Star Mergers can be sources of Axion-Like Particles potentially observable in gamma-rays within the $\sim$10-500~MeV range.
\par We find that a NSM in the nearby galaxies LMC and M33 --admittedly not likely given the known merger rates and the small volume entailed-- could lead to a signal observable by the gamma-ray detector Fermi-LAT, within the ALP parameter space still allowed by present constraints. \par We leave it to future studies to fully explore the ALP parameter space, and to address the strategy and potential of a joint detection of a NSM in both gravitational waves and gamma--rays. Here we limit ourselves to first notice and soundly motivate this concrete possibility, and to point to its potentially ground--breaking implications for astrophysics of extreme environments and physics beyond the standard model.


\par{\bf Acknowledgments}
We are grateful to Pierluca Carenza, Marco Chianese, Alessandro Mirizzi, Giuseppe Lucente, and Pasquale D. Serpico for useful comments. 

This work has been supported by the Italian grant 2017W4HA7S “NAT-NET: Neutrino and Astroparticle Theory Network” (PRIN 2017) funded by the Italian Ministero dell’Istruzione, dell’Università e della Ricerca (MIUR), and Iniziativa Specifica TAsP of INFN. The work of DFGF is partially supported by the {\sc Villum Fonden} under project no.~29388.   This project has received funding from the European Union's Horizon 2020 research and innovation program under the Marie Sklodowska-Curie grant agreement No.~847523 ‘INTERACTIONS’.

\bibliographystyle{unsrt}
\bibliography{references}
\end{document}